\documentclass[fleqn,12pt]{wlscirep}
\usepackage[utf8]{inputenc}
\usepackage[T1]{fontenc}
\usepackage[numbers,super]{natbib}

\usepackage{lineno}

\newcommand{\farcs}{\mbox{\ensuremath{.\!\!^{\prime\prime}}}}
\newcommand{\araa}{Annu. Rev. Astron. Astrophys.}   
\newcommand{\apj}{Astrophys. J.}   
\newcommand{\apjl}{Astrophys. J. Lett.}   
\newcommand{\apjs}{Astrophys. J. Suppl. Ser.}   
\newcommand{\aap}{Astron. Astrophys.}   
\newcommand{\mnras}{Mon. Not. R. Astron. Soc.}   

\newcommand{\nat}{Nature} 

\newcommand{\pasp}{Publ. Astron. Soc. Pacific}   

\newcommand{\galight}{\texttt{galight}}
\newcommand{\sersic}{S\'ersic}
\newcommand{\reff}{R$_{\rm eff}$}

\newcommand{\arcsec}{\mbox{\ensuremath{^{\prime\prime}}}}

\title{Extended Components of Little Red Dots in the Rest-Frame Optical}

\author[1]{Yiyang Zhang}
\author[1,*]{Xuheng Ding}
\author[2,3,*]{Lilan Yang}
\author[4]{Erini Lambrides}
\author[5]{Hollis Akins}
\author[6,7]{Andrew J. Battisti}
\author[8,9]{Caitlin M. Casey}
\author[10,11]{Chang-hao Chen}
\author[3]{Isa Cox}
\author[12]{Andreas Faisst}
\author[13]{Maximilien Franco}
\author[14]{Aryana Haghjoo}
\author[10,11]{Luis C. Ho}
\author[10]{Kohei Inayoshi}
\author[9,15]{Shuowen Jin}
\author[16]{Mitchell Karmen}
\author[17]{Anton M. Koekemoer}
\author[3]{Jeyhan S. Kartaltepe}
\author[1]{Kai Liao}
\author[18,19]{Ghassem Gozaliasl}
\author[20,21]{Masafusa Onoue}
\author[5]{Vasily Kokorev}
\author[22,23]{Namrata Roy}
\author[24]{R. Michael Rich}
\author[22,25,26,27]{John D. Silverman}
\author[25,27]{Takumi S. Tanaka}
\author[1]{Bei You}
\author[28]{Hassen M. Yesuf}
\author[29]{Jorge A. Zavala}

\affil[1]{School of Physics and Technology, Wuhan University, Wuhan 430072, China}
\affil[2]{Department of Physics, School of Physics and Electronics, Hunan Normal University, Changsha 410081, People’s Republic of China}
\affil[3]{Laboratory for Multiwavelength Astrophysics, School of Physics and Astronomy, Rochester Institute of Technology, 84 Lomb Memorial Drive, Rochester, NY 14623, USA}
\affil[4]{NASA-Goddard Space Flight Center, Code 662, Greenbelt, MD, 20771, USA}
\affil[5]{Department of Astronomy, The University of Texas at Austin, Austin, TX 78712, USA}
\affil[6]{International Centre for Radio Astronomy Research, University of Western Australia, 35 Stirling Hwy, Crawley, WA 6009, Australia}
\affil[7]{Research School of Astronomy and Astrophysics, Australian National University, Cotter Road, Weston Creek, ACT 2611, Australia}
\affil[8]{Department of Physics, University of California, Santa Barbara, Santa Barbara, CA 93106, USA}
\affil[9]{Cosmic Dawn Center (DAWN), Denmark}
\affil[10]{Kavli Institute for Astronomy and Astrophysics, Peking University, Beijing 100871, China}
\affil[11]{Department of Astronomy, School of Physics, Peking University, Beijing 100871, China}
\affil[12]{Caltech/IPAC, MS 314-6, 1200 E. California Blvd. Pasadena, CA 91125, USA}
\affil[13]{Université Paris-Saclay, Université Paris Cité, CEA, CNRS, AIM, 91191 Gif-sur-Yvette, France}
\affil[14]{Department of Physics and Astronomy, University of California Riverside, Riverside, CA 92521, USA}
\affil[15]{DTU Space, Technical University of Denmark, Elektrovej 327, 2800 Kgs. Lyngby, Denmark}
\affil[16]{Department of Physics and Astronomy, Johns Hopkins University, Baltimore, MD 21218, USA}
\affil[17]{Space Telescope Science Institute, 3700 San Martin Drive, Baltimore, MD 21218, USA}
\affil[18]{Department of Computer Science, Aalto University, P.O. Box 15400, FI-00076 Espoo, Finland}
\affil[19]{Department of Physics, University of Helsinki, P.O. Box 64, FI-00014 Helsinki, Finland}
\affil[20]{Waseda Institute for Advanced Study (WIAS), Waseda University, 1-21-1, Nishi-Waseda, Shinjuku, Tokyo 169-0051, Japan}
\affil[21]{Kavli Institute for the Physics and Mathematics of the Universe (Kavli IPMU, WPI), UTIAS, Tokyo Institutes for Advanced Study, University of Tokyo, Chiba, 277-8583, Japan}
\affil[22]{Center for Astrophysical Sciences, Department of Physics and Astronomy, Johns Hopkins University, Baltimore, MD 21218, USA}
\affil[23]{School of Earth and Space Exploration, Arizona State University, Tempe, AZ 85287, USA}
\affil[24]{Department of Physics and Astronomy, UCLA, PAB 430 Portola Plaza, Los Angeles, CA 90095-1547, USA}
\affil[25]{Kavli Institute for the Physics and Mathematics of the Universe (WPI), The University of Tokyo, Kashiwa, Chiba 277-8583, Japan}
\affil[26]{Center for Data-Driven Discovery, Kavli IPMU (WPI), UTIAS, The University of Tokyo, Kashiwa, Chiba 277-8583, Japan}
\affil[27]{Department of Astronomy, Graduate School of Science, The University of Tokyo, 7-3-1 Hongo, Bunkyo, Tokyo 113-0033, Japan}
\affil[28]{Key Laboratory for Research in Galaxies and Cosmology, Shanghai Astronomical Observatory, Chinese Academy of Sciences, 80 Nandan Road, Shanghai 200030, China}
\affil[29]{University of Massachusetts Amherst, 710 North Pleasant Street, Amherst, MA 01003-9305, USA}

\affil[*]{e-mail: dingxh@whu.edu.cn, yanglilan@hunnu.edu.cn}

\makeatletter
\renewcommand{\@maketitle}{%
{%
\thispagestyle{empty}%
\vskip-36pt%
{\raggedright\sffamily\bfseries\fontsize{20}{25}\selectfont \@title\par}%
\vskip10pt
{\raggedright\sffamily\fontsize{12}{16}\selectfont  \@author\par}
\vskip25pt%
}%
}%
\makeatother

\begin{document}

\flushbottom
\maketitle
\noindent 
\textbf{Recent JWST observations have revealed a population of red, compact, high-redshift objects called “Little Red Dots” (LRDs), whose host components have remained largely unconstrained, possibly due to their extreme compactness.
Current morphological studies suggest the presence of extended emission in LRDs at rest-frame ultraviolet wavelengths. However, in the rest-frame optical regime, investigations have been limited by small sample sizes and insufficient imaging depth, hindering reliable separation between point-like and potential extended components.
Here we perform the image stacking analysis of 217 LRDs in four NIRCam bands, a large and homogeneous sample observed with the COSMOS-Web survey.
Our results reveal the detection of faint extended emission in the F444W band, with a typical size of $\sim$200 parsecs and magnitude of $\sim$27.7 AB at $z \sim 6.5$.
We perform four-band photometric spectral energy distribution fitting based on galaxy templates and derive an average stellar mass of $10^{9.02^{+0.20}_{-0.18}}$ M$_\odot$.
Given this stellar mass, the host galaxy is compact, that is, $\sim$2.5 times smaller than star-forming galaxies of similar mass at comparable redshifts. This work provides direct observational evidence for the existence of LRD host galaxies at rest-frame optical wavelengths and offers new insights into the stellar buildup of these systems within the first billion years after the Big Bang.}

The discovery of very compact in rest-frame optical, red sources with a characteristic V-shape spectral energy distribution (SED) at high redshift, commonly referred to as Little Red Dots (LRDs), has opened a new challenging into the assembly of the earliest black holes and their environments\cite{Furtak et al.(2023), Greene et al.(2024), Matthee et al.(2024)}. 
Current theoretical interpretations of LRD remain uncertain; therefore, determining the presence and properties of their host galaxies is crucial for constraining their nature and understanding the co-evolution of galaxies and black holes in the early universe.
Recent morphological studies have demonstrated the presence of extended emission in LRDs at rest-frame ultraviolet (UV) wavelengths\cite{Rinaldi et al.(2025)UV, Billand et al.(2025), Chen et al.(2025)}. However, the existence of extended structures (i.e., host galaxy) in the rest-frame optical remains largely unconstrained. Although preliminary attempts have been made for some individual systems\cite{Chen et al.(2025), Rinaldi et al.(2025),  Lin et al.(2025),Chen et al.(2024),Tanakadual et al.(2025)}, a direct and systematic confirmation of extended optical hosts at the population level is still absent.
As the rest-frame optical emission more directly traces the underlying stellar mass, detecting spatially extended emission in this regime will offer particularly strong evidence for the presence of their host galaxies.

We start with a parent sample of 434 LRDs, spanning a redshift range of $z \sim 5$--$9$, originally identified in the COSMOS-Web field~\cite{Akins et al.(2024)}. 
The COSMOS-Web program provides deep imaging with JWST/NIRCam and MIRI over a uniquely large survey area, enabling the construction of a statistical sample of LRD. 
In this study, we utilize NIRCam data in four broadband filters—F115W ($1.15\,\mu$m), F150W ($1.50\,\mu$m), F277W ($2.78\,\mu$m), and F444W ($4.44\,\mu$m),
with $5\sigma$ depth in $0\farcs15$ radius aperture ranges from 26.7 to 27.5 AB mag in F115W and 27.5 to 28.2 AB mag in F444W, depending on the number of integrations\cite{Casey et al.(2023)}.
These NIRCam filters allow us to probe the rest-frame UV and optical emission at $z>5$.
We concentrate on NIRCam data because many LRDs lack corresponding MIRI coverage.

To refine our sample for analysis, we apply three selection criteria to the parent sample: a color cut (F115W $-$ F150W $< 0.8$) to select sources with prominent UV-excess features\cite{Greene et al.(2024),Labbe et al.(2025)}, a magnitude limit ($\mathrm{mag}_{\mathrm{total}} < 30$) to ensure sufficient signal-to-noise ratio (SNR), and a requirement of good imaging fit quality (reduced $\chi^2 < 2$; see Methods). 
In addition, sources showing clear signs of mergers or strong morphological disturbances were excluded through visual inspection, as such systems can introduce substantial uncertainties in the two-dimensional decomposition.
Our final sample comprises 217 LRDs across all four NIRCam bands, though we note that the absence of spectroscopic confirmation for most sources means a certain level of source multiplicity within the sample cannot be excluded, which may introduce additional uncertainty into the population-averaged results.

In our image analysis, we adopt the NIRCam mosaics with a pixel scale of $0\farcs03$, and the details of the imaging data reduction are summarized in Franco et al.\cite{Franco et al.(2025)}.
The fitting is conducted on cutouts of $31 \times 31$ pixels for the short-wavelength bands (F115W, F150W) and $61 \times 61$ pixels for the long-wavelength bands (F277W, F444W), respectively, to account for their different spatial resolutions (see Methods).
During the fitting process, we use Point Spread Function (PSF) models generated via \textsc{PSFEx}\cite{Bertin(2011)}, which models spatial PSF variations across the field as a polynomial function of pixel coordinates by fitting multiple stars simultaneously.  
We adopt \texttt{galight} (v0.2.1) software package \cite{Ding2020} for two-dimensional image analysis.

The observed morphologies of LRDs are typically compact and dominated by the point source\cite{Chen et al.(2024),Furtak et al.(2024)}, 
and the compactness criterion was also adopted to define our parent LRD sample \cite{Akins et al.(2024)}, which requires that $\gtrsim$50\% of the F444W broadband flux be enclosed within a $0.^{\prime\prime}$2 diameter aperture.
Consistent with this, our morphological decomposition shows that in F444W the extended component contributes less than $\sim$30\% of the total flux for the vast majority of sources (see Extended Data Figure~1).
It is challenging to recover the extended emission of LRDs through individual fitting, particularly in the rest-frame optical, where the emission is strongly dominated by the central point source\cite{Matthee et al.(2024)}.
Given that the presence of extended emission in LRDs, particularly in the F444W band, remains uncertain, we first model each source using a single point source (PS-only) across all four NIRCam bands, as shown in the top panel of Figure 1. If any potential extended component exists, we expect to observe central over-subtraction and excess emission in the outskirts within the residual map, that is, signatures of extended light not accounted for by the point-source model.
However, there is no evident emission left in the residual map for an individual LRD as highlighted in the third column, which is probably a result of low SNR.

To enhance the SNR, we stack the residual images of all selected LRDs, obtained after subtracting the point source model and any nearby sources, as shown in the bottom panel of Figure 1. This stacking procedure increases the effective SNR, making faint extended emission more detectable. 
Remarkably, we observe a clear excess of `ring'-like structure in the outskirts, indicating the emission cannot be fully captured by a single point source.
To further verify this detection, we also constructed its corresponding SNR map as presented in the bottom-right panel of Figure 1.
Specifically, we constructed an error map by combining the background noise with the standard deviation of PSFs sampled from 20 different COSMOS-Web tiles.
The total SNR of `ring' structure reaches 72.2, establishing the statistical significance of the extended emission and providing compelling evidence for its existence.
In addition, to rule out the possibility that the excess is caused by fitting uncertainties such as PSF mismatch, we have also performed a control test by using stars, that is, stacking the residuals of 263 stars from PS-only fitting procedure.
We verify that the `ring' structure seen in the stacked residual images of LRD modeling is not an artifact of PSF mismatch (see Extended Data Figure~2).

Having confirmed the presence of extended components, we carry out two-dimensional decomposition for all LRDs using the PS+\sersic\ profile across all four NIRCam bands to characterize the properties of their extended emission \cite{Killi et al.(2024)}. 
To prevent unphysical solutions for the \sersic\ parameters\cite{Sersic1963}, we set the boundaries as follows: effective radius \reff\ $ \in [0\farcs03, 0\farcs1]$ (100–600 pc at z $\sim$ 6.45), \sersic\ index $n \in [1,4]$, and eccentricity $e_1,e_2 \in [-0.3, 0.3]$. The \sersic\ centroid was allowed to vary within two pixels to account for small positional shifts.  
For any nearby contaminants, we model them by additional \sersic\ components.
Based on the results of the PS+\sersic\ fitting described above, we measure the flux of the extended emission and obtain its ratio to total flux.
The distribution of flux ratio of all LRDs across the four NIRCam bands is shown in the left panel of Extended Data Figure~1.
The result shows that the median values of the flux ratio decreases systematically from $\sim$60\% in F115W to $\sim$12\% in F444W,
indicating the extended emission contributes substantially in the blue bands ($<2\mu m$), whereas the active galactic nuclei (AGN) emission dominates in the red bands ($>2\mu m$)\cite{Brooks et al.(2024)}.

To robustly characterize the properties of the extended emission, we perform another stacking analysis based on the PS+\sersic\ decomposition. Specifically, we stack the residual images (data$-$PS$-$nearby sources) of 217 LRDs across four NIRCam bands, as shown in the top panels of Figure 2. During the stacking process, the residual images are re-aligned using the pixel position of the inferred \sersic\ centroid as the reference before stacking, ensuring accurate registration of the extended emission (see Methods).
We then fit the stacked data via a single \sersic\ model (see Figure 2, middle panels).
The best-fit \sersic\ parameters (magnitude, \reff, $n$), and the flux ratio between extended emission and total flux, are summarized in Table~\ref{tab:host_props}, which are considered as representative of the average extended emission properties.
The averaged magnitude of the extended component is close to 29 mag, which is well below the individual detection threshold of the COSMOS-Web 5$\sigma$ detection limits (27.4--28.2 AB mag).
We emphasize that though a subset of LRDs may exhibit individually detectable extended structures, our stacking approach is designed to characterize the population-average properties of uniformly selected sources. By combining a large sample under consistent selection criteria, the stacked results reflect the typical behavior of the LRD as a population, rather than being biased by a small number of morphologically complex objects, particularly given that residual broadening arising from the cumulative effect of small positional offsets during stacking cannot be completely excluded.
Simulation tests further confirm that the stacking and fitting procedure can reliably recover the input extended emission parameters to within 0.4 mag in magnitude and 0.2 dex in size (see Extended Data Figure~3).

The measured effective radius \reff\ of the stacked data is $\sim$210 pc in the rest-frame optical (F444W) and increases toward shorter wavelength, i.e., $\sim$392 pc in the rest-frame UV (F115W).
Our LRDs often exhibit more complex and heterogeneous morphologies in the rest-frame UV bands, such as clumpy or multicomponent structures, which is consistent with previous UV-based studies, e.g., Rinaldi et al.\cite{Rinaldi et al.(2025)UV}; Chen et al.\cite{Chen et al.(2025)}.
This probably contributes to the larger effective sizes measured in the shorter wavelength. We note, however, that these irregular morphologies may also influence the measured average extended component, and therefore should be considered when interpreting the wavelength dependence of the inferred sizes.
The wavelength dependence suggests that the UV emission, tracing recent star formation, is more spatially extended than the underlying evolved stellar distribution probed by optical light. 
Such extended UV components are consistent with the radial-profile evolution reported by Rinaldi et al.\cite{Rinaldi et al.(2025)UV,Rinaldi et al.(2025)}, as well as similar findings in Billand et al.\cite{Billand et al.(2025)} and Chen et al.\cite{Chen et al.(2025)}.
The flux ratios of the stacked extended-to-total flux are approximately 40\% and 10--20\% in the shorter and longer wavelengths, respectively.
The notable UV contribution from the extended emission is consistent with the BH–star model hypothesis \cite{Naidu et al.(2025)}, which suggests that the rest-frame UV part of the V-shaped SED mainly originates from the galaxy, whereas the overall V-shaped SED arising from the combination of galaxy and BH–star contributions.

We utilize the inferred four-band photometry of the stacked source and perform spectral energy distribution (SED) fitting to clarify its physical origin. 
We use \texttt{bagpipes}\cite{bagpipes} to perform SED fitting.
We first adopt a single combined stellar population model at an average redshift of $z=6.45$ to interpret the extended emission as originating from stars.
The stellar metallicity is allowed to vary between $0$ and $2.5~Z_\odot$, the stellar age is constrained to $[0.01, 0.85]$~Gyr (capped by the age of the Universe at the source redshift), and the dust attenuation is modeled with a Calzetti law, with $A_V$ (dust attenuation in the V band) free to vary in the range $[0, 5]$~mag. Our result suggests that the stacked host of LRDs is consistent with a star-forming galaxy (SFG) with a stellar mass $M_*$ of $10^{9.02^{+0.20}_{-0.18}} M_{\odot}$ (see Figure 3).
The inferred properties of the host match with the expected mass range $10^{8.5}$ to $10^{10} M_{\odot}$ of star-formation-dominated LRDs as reported in Akins et al.~\cite{Akins et al.(2024)}, and also agree with the $\sim 10^{9} M_{\odot}$ stellar mass of an unusually bright dust-reddened AGN host galaxy reported by Wang et al.\cite{Wang et al.(2024)}.
We have also verified the robustness of our results by cross-checking them with other SED fitting codes, such as \texttt{gsf} \cite{gsf}, yielding consistent outcomes.

Because our stacking analysis combines LRDs across a broad redshift range, we test whether this redshift spread could bias the inferred host properties. We therefore repeat the analysis by dividing the LRDs into finer redshift bins with width of $\Delta z = 0.5$.
We find that the inferred host sizes and magnitudes remain broadly consistent across redshift bins, showing no significant evolution within the uncertainties.
For the samples at $z \sim 7\text{--}9$, the blue excess is slightly enhanced, which could potentially arise from additional contributions such as $\mathrm{Ly}\alpha$ emission.
The sizes in F444W increase mildly from $\sim$212 pc at $\langle z \rangle = 5.78$ to $\sim$255 pc at $\langle z \rangle = 7.15$, and the corresponding stellar masses range from $\log(M_*/M_\odot) \approx 8.80$ to $9.55$.
The slightly higher stellar masses inferred at higher redshift are probably driven by selection effects, as the apparent magnitude distributions of LRDs across redshift bins are similar. This implies that intrinsically more luminous and, hence, more massive sources are preferentially included at higher redshift. Overall, these tests indicate that our main results are not driven by redshift mixing and remain robust across the probed redshift range.

Our SED fitting results suggest negligible dust attenuation in the extended emission, with a low inferred $A_V$, although some theoretical models predicted that LRDs should exhibit strong dust emission\cite{Li et al.(2025),Delvecchio et al.(2025)}.
Moreover, observations from ALMA and JWST/MIRI studies have explored the mid- and far-infrared properties of LRDs, with some results suggesting limited dust content in these sources \cite{Setton et al.(2025), Akins et al.(2025), Xiao et al.(2025)}.
These initial findings indicate low dust content, but the current constraints are still limited by small sky coverage and relatively shallow depth \cite{Perez-Gonzalez et al.(2024)}.
Moreover, the near-infrared slopes span a broad range of values, indicating that the available data are not sufficient to robustly reveal the truth of dust. Therefore, the definitive conclusion of dust properties needs to be further contained by more data in the future.
Furthermore, according to the empirical relations between stellar and dust mass, given their relatively low stellar masses, LRDs host are expected to have dust masses below $10^{6}~M_{\odot}$ \cite{Donevski et al.(2020),Casey et al.(2025)}.
Such low dust masses probably result in dust emission levels below current ALMA detection limits, providing a natural explanation for the lack of significant detection.

We also explored alternative SED interpretations by fitting the stacked photometry using a nebular gas model and a composite model that combines stellar and gas components. The gas-only model can reproduce the UV flux under certain fitting assumptions (see Methods). However, it cannot adequately reproduce the rest-frame optical flux in F444W. In addition, it requires strong emission-line contributions which are not supported by current spectroscopic constraints\cite{Hviding et al.(2025)} (see Extended Data Figure~4). When allowing both stellar and gas components to vary freely, the composite model naturally converges towards a stellar-dominated solution, with only a minor gas contribution unless an explicit lower bound is imposed (see Methods). 
The above analysis indicates that the strong nebular contribution is unnecessary to explain the extended emission shown in the stacked photometry, particularly in the rest-frame optical regime where the constraints are most robust. The V-shaped SED of the extended component may be partially driven by residual systematics in the low-SNR blue bands, whereas the apparent excess at the blue end could also include contributions from diffuse gas emission that is difficult to quantify. Our result indicates that the extended emission is primarily stellar-dominated, with a possible minor contribution from gas at short wavelengths. It is worth noting that the separated point-source components also exhibit a V-shaped SED, suggesting that this feature is not uniquely associated with either the central source or the extended host component.

The host galaxy is compact in the rest-frame optical with $R_{\rm eff}=210^{+179}_{-95}$ pc at an average redshift of $z=6.45$.
Adopting the measured stellar mass from SED via stellar template, we compare our data point highlighted by the red star to star-forming galaxies (SFGs) and quiescent galaxies (QGs) in the size--mass plane at $6<z<7$.  The results of SFGs and QGs are also obtained from the COSMOS-Web field\cite{Yang et al.(2025)}.
We also demonstrate the results of LRD hosts obtained from divided redshift bins.
As shown in Figure 4, LRD hosts occupy the low-mass end of the distribution, with sizes among the smallest observed at these stellar masses. In the mass range of $ \log(M_*/M_\odot) = 8.6\text{--}9.0 $, the median size of SFGs is $\sim 518$ pc, approximately 2.5 times larger than that of LRD hosts.  
The extremely compact nature of LRD hosts suggests that they have accumulated a high density of central gas and experienced concentrated star formation, thus assembled a substantial stellar mass in central region. The buildup of the dense gas in the core would naturally fuel the central supermassive black hole, triggering the ignition of AGN\cite{Inayoshi et al.(2025)}.
Contrast to the compact nature in the rest-frame optical,  the size measured in UV is extended, i.e., $\sim 400$ pc. 
According to SED analysis, the extended UV emission is dominated by a young stellar population with low dust attenuation, albeit it could also possibly be from strong nebular emissions.
In the stellar component dominated scenario, the outskirts of the host are undergoing recent star formation, resulting in a larger size in UV than in optical. This provides evidence for the early stage of inside-out galaxy growth.

Assuming that the local scaling relations between supermassive black holes and their host galaxies hold at $z\sim6$ \cite{KH13,G20}, the inferred host stellar masses of $\log(M_*/M_\odot)=9$ imply black hole masses of $\log(M_{\rm BH}/M_\odot)=6\text{--}7$.
This inferred mass range of black hole is broadly consistent with several recent theoretical interpretations of LRDs, including gas-enshrouded black hole scenarios\cite{Inayoshi et al.(2025),Umeda et al.(2025)}.
However, this mass range is generally lower than some estimates derived from broad emission-line measurements in LRDs, highlighting the current uncertainties in black hole mass determinations for this population.

Our detection of a resolved rest-frame optical component demonstrates that LRDs host substantial stellar components, indicating that they are not purely point-like systems. Regardless of the detailed physical interpretation of LRDs, our results imply that these sources reside within compact host galaxies, providing an important observational constraint for theoretical models, which must account for the presence of a stellar host surrounding the central black hole.

\clearpage
\newpage

\section*{Tables}
\begin{table}[ht]
    \centering
    \begin{tabular}{lccccc}
    \hline \hline
   Filter & Magnitude & $\log$\reff ($arcsec$) & $\log$\reff ($pc$) &\sersic\ index ($n$) & Flux ratio \\
    
    \hline
    F115W & $28.91\pm0.44$ & $-1.16\pm0.37$ & $2.59\pm0.37$ & $1.66\pm0.70$ & $\sim39.6\%$ \\

    F150W & $29.23\pm0.55$ & $-1.21\pm0.42$ & $2.54\pm 0.42$ & $2.00\pm1.04$ & $\sim33.9\%$ \\

    F277W & $28.76\pm0.19$ & $-1.37\pm0.18$ & $2.38\pm 0.18$ & $1.00\pm0.14$ & $\sim23.2\%$ \\

    F444W & $27.71\pm0.13$ & $-1.42\pm0.26$ & $2.33\pm 0.26$ & $1.00\pm0.12$ & $\sim11.8\%$ \\
    \hline \hline
    \end{tabular}
    \caption{{\bf Properties of stacked extended emission across four NIRCam bands.}
    The columns from left to right represent the filter, magnitude, half-light radius, S\'ersic index and flux ratio, respectively.
    The $\log$\reff ($pc$) are calculated assuming a redshift of $z = 6.45$, and
    flux ratio is the ratio between the stacked extended emission and the averaged total flux, see more details in Extended Data Figure~1.
    All uncertainties are derived from simulation and bootstrap analysis (see Methods).}
    \label{tab:host_props}
\end{table}

\newpage

\section*{Figures}
\begin{figure}[h]
\centering
\includegraphics[trim = 0mm 0mm 0mm 0mm, clip, width=\textwidth]{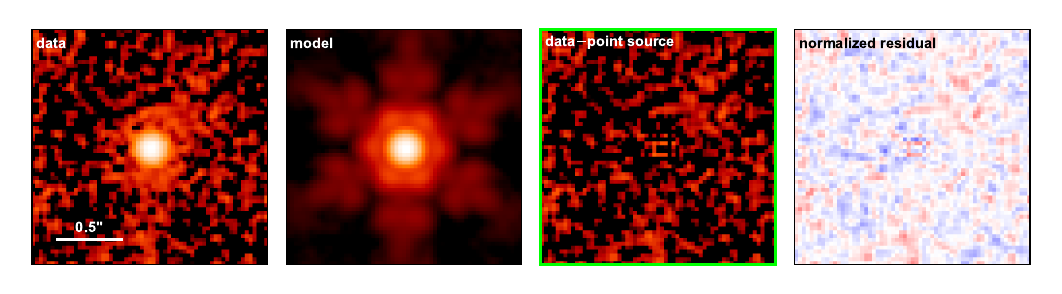}
\includegraphics[trim = 0mm 0mm 0mm 0mm, clip, width=\textwidth]{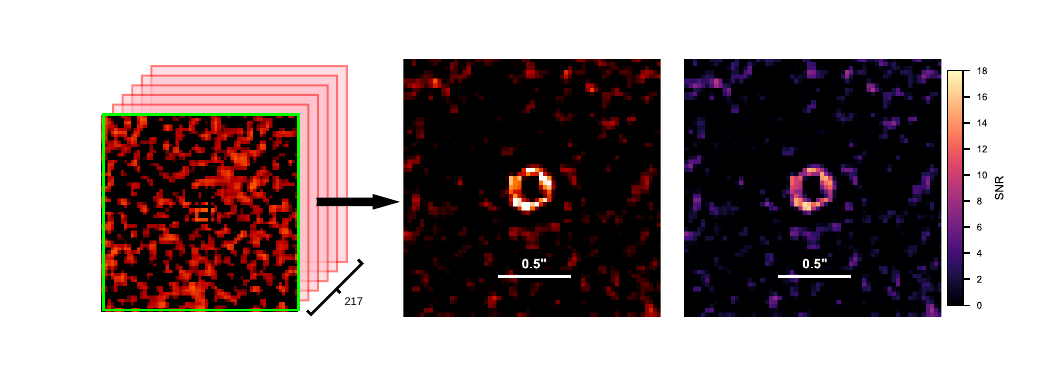}
\caption{{\bf Stacking approach reveals extended emission in LRDs at F444W.}
{\it Top:} An example of 2D decomposition for an individual LRD using the PS-only model. From left to right columns: original LRD image (data), model image generated from the PSF (model), residual image after subtracting the point source component (data$-$PS), and the normalized residual.
{\it Bottom left:} Illustration of the imaging stacking process.  Residuals (data$-$PS) from 217 LRDs are aligned and averaged to produce a stacked image that enhances the visibility of low-surface-brightness extended emission in the outskirts.
{\it Bottom right:} 
Stacked data$-$PS image and its  SNR map. 
The standard deviation image (error map) is computed by measuring pixel-by-pixel dispersion across different PSFs at the same location and corresponding background noise, and is used to perform SNR calculation.
}
\label{fig:PSonlystack}
\end{figure}

\begin{figure}
\centering
\includegraphics[trim = 0mm 0mm 0mm 0mm, clip, width=\linewidth]{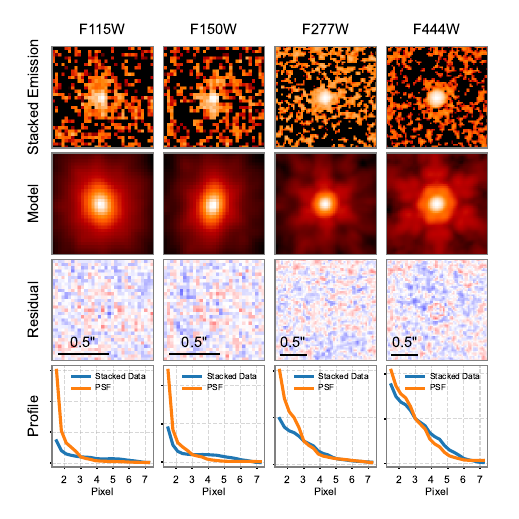} \\
\caption{
{\bf Image modeling of the stacked extended emission across four NIRCam bands.} 
{\it Top:} Stacked emission images (data$-$PS) constructed by stacking the PS-subtracted images of 217 individual LRDs that are modeled via PS+\sersic\ decomposition across four NIRCam bands.
This illustrates the wavelength-dependent morphology of the mean extended component. 
A scale bar of 0.5$\arcsec$ corresponds to $2.8$~kpc at $z\sim6.45$. 
{\it Second row:} Best-fit models using a single \sersic\ model. 
{\it Third row:} Corresponding normalized residuals.
{\it Bottom:} Azimuthally averaged and normalized light profiles of the stacked data (blue) with comparison to the PSF (orange), and the profiles are normalized at a radius of 3 pixels.}

\label{fig:StackFit}
\end{figure}

\begin{figure}
\centering
\hspace{-1.7cm}
\includegraphics[trim = 0mm 0mm 0mm 0mm, clip, width=\linewidth]{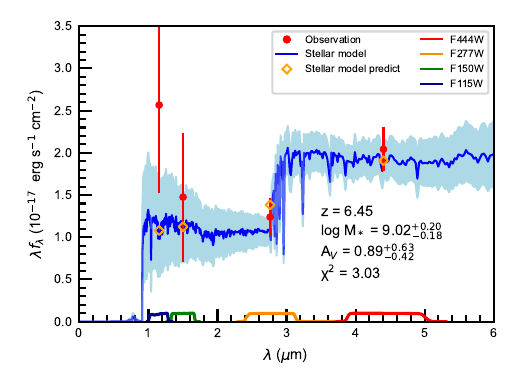} \\

\caption{{\bf SED fitting of the host based on four-band photometry using a stellar population template.}
The red data points show the inferred host fluxes with error bars demonstrating the 1$\sigma$ error.
The blue line represents the best-fit stellar template, with shaded region indicating the 1$\sigma$ uncertainty derived from Markov-Chain Monte-Carlo (MCMC) sampling.
The orange diamonds indicate the predicted fluxes from the stellar template across the four NIRCam filters.
Comparing to the observations (red points), there is an excess at short wavelengths (notably F115W), which probably implies
additional blue-end contributions beyond a simple stellar population model.
}
\label{fig:SED}
\end{figure}

\begin{figure}
\centering
\hspace{-1.7cm}
\includegraphics[width=\linewidth]{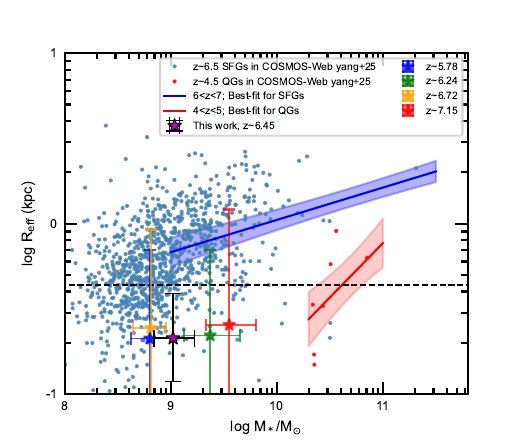} 
\caption{
{\bf LRD hosts on the size–mass plane.}
The magenta star highlights the averaged result derived from the full LRD sample at $z\sim6.45$, with the uncertainty in size estimated from our simulation-based error analysis and the stellar mass uncertainty derived from SED fitting. While the other stars in blue, green, yellow, and red represent measurements in divided redshift bins spanning $z = 5.5\sim7.5$ with a bin width of $\Delta z = 0.5$. For these measurements, the uncertainties in size rescaled according to the number of stacked sources in each redshift bin, while the stellar mass uncertainties are obtained from SED fitting.
The blue and red points represent star-forming galaxies (SFGs) at $z \sim 6.5$ and quiescent galaxies (QGs) at $z \sim 4.5$, respectively, which are also obtained in the COSMOS-Web field\cite{Yang et al.(2025)}. 
The blue and red lines show the best-fit size--mass relations for SFGs with $M_* > 10^9\, M_\odot$ and QGs.
LRD hosts exhibit systematically smaller sizes compared to other star-forming galaxies of similar stellar mass, suggesting that their structural growth may still be in an early stage.
The horizontal dashed line denotes the PSF size of NIRCam/F444W at redshift 6.5.
}
\label{fig:size-mass}
\end{figure}

\clearpage
\newpage
\noindent \textbf{\Large Methods}

\medskip
\noindent {\bf Target selection and data imaging.} \label{sec:Target selection and data imaging}
Our dataset is from the COSMOS-Web survey, the largest NIRCam imaging survey covering $0.54~\mathrm{deg}^2$ in the JWST Cycle 1 treasury program.
The COSMOS-Web survey\cite{Casey et al.(2023),Casey et al.(2024)} (GO:1727, PI: J. Kartaltepe and C. Casey) covers a part of the COSMOS field \cite{Scoville et al.(2007)} with NIRCam\cite{Rieke et al.(2023)} in four filters (that is, F115W, F150W, F277W, and F444W) with approximate $5\sigma$ depths ranging from 27.4 to 28.2 AB mag\cite{Casey et al.(2024)}.
Our parent LRD sample consists of 434 sources identified by Akins et al. \cite{Akins et al.(2024)}, with redshift spanning $z\sim5$--$9$. The photometric redshift and magnitude distributions are shown in Supplementary Figure~1. The robustness of COSMOS-Web photometric redshifts has been extensively validated through large spectroscopic comparisons in the literature \cite{Huberty et al.(2026), Khostovan et al.(2026)}.
Those sources were selected based on a compactness criterion, $C_{444} = f_{444}(d = 0.2'')/f_{444}(d = 0.5'') > 0.5$, where $f_{444}$ is the F444W-band flux and $d$ is the diameter aperture, meaning that $\gtrsim$50\% of the F444W broadband flux measured within a 0\farcs5 diameter aperture is enclosed within a 0\farcs2 diameter aperture, and a red color criterion of $F277W - F444W > 1.5$, designed to isolate compact sources with distinctly red colors. This selection naturally favors systems whose F444W emission is centrally concentrated, and therefore excludes sources with strongly irregular or disturbed morphologies in the F444W band.

In this work, to capture prominent UV-excess features and to ensure robust photometric measurement, we further refine the sample by applying three additional selection criteria: (1) a blue color cut of $F115W - F150W < 0.8$; (2) a total magnitude cut $\mathrm{mag}_{\mathrm{total}} < 30$
to guarantee sufficient SNR for reliable analysis; and (3) a goodness-of-fit threshold with reduced $\chi^2 < 2$ based on PS+\sersic\ fitting (see next two subsections for details). 
The magnitude cut serves as a practical criterion to ensure sufficient photometric quality rather than a strict signal-to-noise threshold. In practice, it removes only a small fraction of sources, with most exclusions driven by the image-fit quality and color criteria.
These criteria reduce our final sample to 217 LRDs, which have high-quality, homogeneous photometric measurements across four NIRCam bands. We also performed a test without applying the first criteria (i.e., color selection), resulting in a sample of 357 LRDs. The derived magnitudes and sizes remain consistent ($<0.4$ mag and $<0.2$ dex, respectively) with those from our main analysis, except for F115W in which the inferred extended component in the short-wavelength bands appears fainter due to the absence of the V-shape selection. Nevertheless, the inferred stellar mass remains robust and consistent across samples.

The data presented in this work are taken with Module A and Module B of the NIRCam instrument, which has a field of view of $2.2\times2.2$ square arcminutes for each module. Images in the four bands (F115W, F150W, F277W and F444W) were drizzled to a pixel scale of $0\farcs{}03$ \cite{Franco et al.(2025),Bushouse et al.(2023),Rest et al.(2023),Bagley et al.(2023),Koekemoer et al.(2007)}.

The COSMOS-Web survey was conducted in three epochs between January 2023 and January 2024. For LRD objects observed multiple times in overlapping regions, we performed the fitting for all the available data. As the measurements across different epochs are generally consistent, we adopt the results from the earliest observation in our analysis.

\medskip
\noindent {\bf 2D image fitting preparation.} \label{sec:2D image fitting preparation}
We carry out 2D image analysis on all parent sample including 434 LRDs from the COSMOS-Web field using the open-source Python package \galight\ software, which utilizes the image modeling capabilities of \texttt{lenstronomy}\cite{2021JOSS....6.3283B}. This tool has been shown to robustly recover total magnitudes, achieving an accuracy better than 0.2 mag for galaxies as faint as 27 mag \cite{Tang et al.(2025)}.
To account for different spatial resolutions across filters at different wavelengths, we adopt different cutout sizes. Specifically, the F444W and F277W images are cropped to $61\times61$ pixels, while the F150W and F115W images are cropped to $31\times31$ pixels.
The global background light subtraction is first performed using the \texttt{Background2D} function of \texttt{Photutils}\cite{Bradley2022}. Since our method involves combining and stacking data from over 200 images, even negligible background light present in individual sources can be amplified and become non-negligible in the final stacked image. To address this, we apply an additional flat background correction to each LRD image to ensure that the background light in the empty region remained statistically neutral after model fitting. This step ensures an unbiased fitting for both the total flux and the $R_{\rm eff}$, defined here as the semi-major-axis half-light radius of the \sersic\ component.\cite{Ding et al.(2021)}.

Since the COSMOS-Web field is divided into multiple subregions (A1–A10 and B1–B10), we derive the PSF models obtained within each subregion to match the local optical response. JWST mirrors are rephased every few weeks, thus, minor PSF variations across epochs may introduce small systematic uncertainties, though these are expected to average out in the stacking process.
The PSF models are constructed via \texttt{PSFEx} \cite{Bertin(2011)},
which builds empirical PSFs from unsaturated stars across the field and models spatial variations as low-order polynomial functions of detector coordinates. 
Because \texttt{PSFEx} derives the PSF shape by combining multiple stellar profiles with different brightness levels, the resulting PSF is not tied to any single flux scale and is therefore robust against potential mismatches between the brightness of PSF stars and the science targets.
These PSF models have also been widely used in many previous COSMOS-Web analyses (e.g., Yang et al.\cite{Yang et al.(2025)}, Tanaka et al.\cite{Tanaka et al.(2024),Tanaka et al.(2025)}, Zhuang et al.\cite{Zhuang2024}).
To further assess the robustness of adopted PSF models, we have also performed an independent validation using PSFs constructed directly from real stars in the field (see Section ``Robustness of the extended emission to alternative star-based PSFs'').

\medskip
\noindent {\bf Fitting strategies.} \label{sec:Fitting strategies}
Two different fitting models were used to fit the data: (1) a PS-only model, where the entire flux distribution is represented by a single PSF component, and (2) a composite model consisting of a PSF component and a single \sersic\ profile convolved with the PSF. In this second model, the \sersic\ model is used to capture the extended component of LRD.

The \sersic\ component includes seven free parameters: amplitude, \sersic\ index $n$, effective radius $R_{\rm eff}$, central coordinates $(x_c, y_c)$, and eccentricity parameters $(e_1, e_2)$. 
To avoid unphysical parameter inference, we applied constraints to the \sersic\ profile parameters.
Regardless of the fitting method (PS-only or PS+\sersic), the initial source coordinates $(x_c, y_c)$ were adopted from Akins' catalog\cite{Akins et al.(2024)}. 
For the PS+\sersic\ fits, we limit the \sersic\ centroid deviation to within 2 pixels of the PSF center to avoid positional shifts due to low SNR and background contamination.  
The effective radius \reff~$\in[0\farcs{}03 ,0\farcs{}1]$ (i.e., from one pixel to up to 600 pc) to prevent overestimation due to noise. Given the intrinsic compactness of LRD, unconstrained fitting often resulted in unrealistically high $n$ values ($n > 8$). We thus set the \sersic\ index ($n)\in[1,4]$, consistent with typical morphological expectations: $n=1$ represents disk-like structures, while $n=4$ corresponds to classical bulges. The eccentricity parameters $e_1$ and $e_2$ were limited to the range $-0.3$ to $0.3$ to maintain physically plausible ellipticities. 
For the LRD with any nearby object in the cutout, we simultaneously fit with another \sersic\ profile with no constraints, allowing us to remove its light during the stacking.

\medskip
\noindent {\bf Results of individual source fitting.} \label{sec:Results of individual source fitting}
We fit each source via two fitting strategies introduced above.
For illustration, we present in Supplementary Figure~2 the fitting results for an example source in the F444W filter. The figure compares the results obtained using the PS-only model (top panel) and the PS+\sersic\ model (bottom panel).
For the same target, the PS+\sersic\ model reveals extended emission, while the PS-only model also yields comparably good residuals. This illustrates the inherent degeneracy in individual fits, where, given the compact nature of LRDs and the limited imaging depth, it is difficult to robustly determine whether the recovered extended component is physically meaningful for single source\cite{Rinaldi et al.(2025)UV,Billand et al.(2025),Pacucci and Loeb(2025)}.
This ambiguity in individual-object fitting motivates our stacking-based approach, which is designed to mitigate noise-driven degeneracies and to assess the presence and properties of extended emission in a statistical sense. Rather than relying on uncertain decompositions of individual LRDs, we use stacking to characterize the average extended emission properties of the LRD population as a whole.
We compare the Bayesian Information Criterion (BIC) of those two fittings, and show no significant difference in goodness-of-fit between the two models; this result holds across the entire parent LRD sample.
This similarity is probably due to the faintness of the extended emission and the limited SNR in individual sources, which constrain the ability of the BIC to distinguish the models.
In the following subsection, we provide compelling evidence for extended components in higher SNR data produced by stacking.

We now examine the significance of the extended emission revealed by PS+\sersic\ modeling and its distribution across four NIRCam filters.
Specifically, for sources with $\chi^2 < 2$ based on PS+\sersic\ fitting, we derived the extended-to-total flux ratio based on the decomposition model, defined as the fraction of flux between the \sersic\ component and total, i.e., PS+\sersic.
The distribution of extended-to-total flux ratio across four bands of the final selected 217 LRDs is shown in Extended Data Figure~1 (left panel). 
The distribution skews toward larger values (70\%--80\%) at F115W and F150W, however, skews to smaller values at F277W and F444W. For example, the distribution at F444W peaks at $\sim10\%$.
The larger fraction of LRD has a higher extended-to-total flux ratio at shorter-wavelength filters, whereas this ratio decreases substantially towards longer-wavelength filters.
To investigate the redshift dependence of the extended-to-total flux ratio, we divided the LRD sample into redshift bins of width $\Delta z = 0.5$, spanning the range $5 < z < 9$. For each of the four \textit{JWST}/NIRCam bands (F115W, F150W, F277W, and F444W), we computed the mean value within each redshift bin and presented in Extended Data Figure~1 (right panel).
Overall, the change as a function of redshift is not significant, although the flux ratio decreases slightly toward the higher redshift. This may be due to the small bin size and the potential redshift bias, which makes it less likely to detect weak extended structures around dense sources at higher redshifts than at lower redshifts.

\medskip
\noindent {\bf Unveils faint extended emission via image stacking.} \label{sec:Unveils faint extended emission via image stacking}
We used a stacking method to enhance the SNR in the residual maps of LRD$-$PS image. Statistically, the signal increases linearly with the number of stacked images $N$, and the noise grows only as $\sqrt{N}$, leading to an overall enhancement $\propto \sqrt{N}$, which is $\sqrt{217}$ according to our sample size. We collected the residual maps obtained by subtracting the best-fit point source from each LRD and stacked these images, aligning them based on the LRD center positions. To minimize contamination, we also subtract flux from neighboring objects within the field of view. This ensures that the final stacked image predominantly contains only background noise and any genuine extended emission. We perform a simple (unweighted) average stack of the residual images of 217 LRDs, according to our target selection. The resulting stacked image is normalized by the number of adopted sources, representing a high-SNR image highlighting any extended structures after removing the PSF model.

We first present the image stacking results for the PS-only model. As shown in Figure 1, we demonstrate the stacking method and SNR map of the stacked residual image at the F444W band.
Our stacked image reveals noticeable over-subtraction in the central region, characterized by negative flux, indicating that the point source profile inferred by the PS-only model overestimates the central flux.
Meanwhile, the residual light in the outskirts forms a positive flux `ring' feature with a typical radius of $\sim$0.2 arcsec,
suggesting the presence of extended emission that cannot be accounted for by a single point source. This outcome is expected when the fitted data contains unresolved extended components that the PS-only model fails to capture. To verify that the `ring' feature is not an artifact caused by PSF mismatch to the data, we calculated the flux ratio of the `ring' feature and compared it to the standard deviation of the PSFs derived from all PSFs in the COSMOS-Web field.
We find that the flux level of the `ring' feature exceeds the PSF uncertainty at the corresponding region, with an SNR of approximately 10.
This result provides strong evidence for the presence of extended emission in the stacked LRD sample, which cannot be explained by PSF uncertainty.

Given the confirmed presence of extended components, we adopt a PS+\sersic\ model to represent the central point source and the underlying extended component.
As shown in Figure 2,
we now perform the residual stacks derived from data$-$PS component obtained from PS+\sersic\ modeling, across all four bands (F115W, F150W, F277W, and F444W) (see Figure 2, top panel). 
This provides a clearer view of the extended emission structure, enabling more detailed subsequent analyses.
We estimate the SNR of the stacked image in each band. 
The measured SNRs are 8.1, 7.0, 15.2, 32.6 in F115W, F150W, F277W, and F444W, respectively, indicating that the extended emission detections are statistically significant, particularly in the longer-wavelength bands, where the extended emission is more pronounced. For comparison, the average SNR of the extended emission measured from individual sources is only 0.5, 0.6, 2.8, and 3.3 in F115W, F150W, F277W, and F444W, respectively, demonstrating the substantial improvement in sensitivity achieved through stacking.

We note that the relative positions between the extended component and the point source are not always centered on individual pixels, which could introduce some dispersion in the stacked image. We therefore examined the relative offsets between the \sersic\ component (representing the extended component) and the point source component in the individual image fits. We find that the largest offsets are of order one pixel and remain within two pixels, indicating that the degree of miscentering is generally small.
In our analysis, the centroid of the inferred extended (\sersic) component is used as the reference for image alignment. Each residual image is shifted to this centroid using direct pixel translation without interpolation before stacking.
Specifically, images are shifted by integer numbers of pixels along both axes, and regions moved outside the original frame are filled with zeros. Given that the characteristic size scale of the recovered extended emission is itself comparable to the pixel scale, this alignment strategy minimizes potential biases in size measurements.

We also estimated the extended-to-total flux ratio from the stacked image, defined as the flux of the extended component divided by the total flux. As shown in Extended Data Figure~1, due to the lower SNR of individual LRD in the shorter-wavelength bands, many sources were misidentified as having high extended-to-total flux ratios during individual fitting, resulting in a notable discrepancy between the stacked ratio and the median of the individual measurements. Based on the stacked extended-to-total flux ratio, the point source component clearly dominates in the longer-wavelength bands, contributing up to 90\% of the total flux. Even in the shorter-wavelength bands, the flux is not entirely from the extended emission, suggesting that additional mechanisms beyond extended emission may contribute to the rising side of the LRD's characteristic V-shape.

\medskip
\noindent {\bf Control tests for the `ring' feature.} \label{sec:Validating the PSF model accuracy using star images.}
To assess whether the `ring'-like signal observed in the PS-only stacked residual image of LRD could be artifacts introduced by the fitting procedure or PSF modeling, we conducted a control test using a star from the same COSMOS-Web field.
We selected 263 isolated, unsaturated stars with F444W magnitudes fainter than 24 mag to ensure a fair comparison with the LRDs in terms of SNR and flux level. Each star was modeled using the same PS-only fitting procedure and the corresponding PSF from \textsc{PSFEx} as applied to the LRD sample. The fitted residual images were stacked following the same stacking procedure described in main text.

The resulting stacked residual image of stars exhibits no significant extended structures or `ring'-like patterns, appearing compact and symmetric as in Extended Data Figure~2. This contrasts with the LRD residual stack and confirms that the residual features in LRD are not due to fitting artifacts or imperfections in the PSF model, but instead reflect real extended emission beyond the point source.

As a further validation, we performed an additional control test using compact galaxies located in the same COSMOS-Web field. We selected 45 compact galaxies whose intrinsic sizes are smaller than half of the PSF in the F444W band, such that they are expected to be largely unresolved or only marginally resolved. These compact galaxies were modeled using the same PS-only fitting procedure and PSF models as applied to the LRDs, and their residual images were stacked following the identical procedure.

Interestingly, the fitting based on PS-only stacked residual image of these compact galaxies exhibits a `ring'-like excess that closely resembles the feature observed in the LRD PS-only residual stack. By contrast, no such structure is present in the stacked residuals of stars. The consistent behavior observed for two physically distinct source populations, i.e., stars and compact galaxies, demonstrates that the `ring' feature arises naturally when an underlying extended component is present. This strongly supports the interpretation that the `ring'-like signal seen in the LRD PS-only stacked residual image originates from genuine extended emission, rather than from PSF modeling uncertainties or fitting artifacts.

\medskip
\noindent {\bf Robustness of the extended emission to alternative star-based PSFs.}
As an additional robustness check on the PSF modeling, we repeated the full fitting analysis using an alternative empirical PSF construction method independent of \textsc{PSFEx}. Specifically, the COSMOS-Web field was divided into 20 subregions (A1--A10 and B1--B10). In each region, we selected the best-quality star in each band, defined as having the smallest FWHM and no detectable contamination from nearby sources. These stars were then stacked to construct 20 independent band-specific empirical PSFs, hereafter referred to as the star-PSF set.

Using these star-based PSFs, we repeated the PS-only fitting and stacking procedure applied in the main analysis. The resulting stacked residual images again exhibit a `ring'-like structure consistent with that obtained using the \textsc{PSFEx} PSF models, demonstrating that the detected extended residual feature is not sensitive to the specific PSF construction method.

We further quantified the impact of PSF choice on the recovered extended emission properties (see next subsection). Compared to the results obtained using the \textsc{PSFEx}, the difference in the recovered extended emission magnitude is below $<0.024$ mag, and the effective radius is below $<0.005$ arcsec. Both differences lie well within the $1\sigma$ measurement uncertainties, indicating that our main conclusions are robust against reasonable variations in the PSF model.

\medskip
\noindent {\bf Photometry and morphology of the stacked extended emission.} \label{sec:Photometry and morphology of the stacked extended emission}
We begin by measuring the apparent sizes of the extended component and comparing them to the corresponding PSFs. 
The full width at half maximum (FWHM) values are obtained by averaging measurements along four directions—horizontal, vertical, and two diagonals—on both the stacked image and the PSFs.
We find that the stacked image exhibits FWHM consistently larger than those of the PSFs, confirming their extended nature. 
Importantly, the measured extended component FWHMs are significantly broader than the PSF sizes. 
As mentioned in the ``Robustness of the extended emission to alternative star-based PSFs'' section, the extended component positions are not precisely at the pixel center, which could introduce a positional dispersion of at sub-pixel scale. However, this level of dispersion cannot account for the observed broadening of the stacked profiles we observed, indicating that the extended sizes are not due to PSF mismatches but rather reflect genuine extended component.

To accurately measure this extended component, we fit a single \sersic\ model to the stacked image. Likewise, the fitting on the individual source, we restrict the \sersic\ index ($n)\in[1,4]$ to ensure physical plausibility, while allowing a wide range of the $R_{\rm eff}$. 
The PSFs used for reconstruction are selected from the B1 region in each band, and we find that the fitting results remain consistent using different PSFs across the field.

We list the fitting results in Table~\ref{tab:host_props}, the error bars of these numbers are given using the combination of the simulation tests and bootstrap method (see next subsection).
The best-fit value for the fitted magnitudes of the extended components is 28.9, 29.2, 28.7, and 27.7 for F115W, F150W, F277W, and F444W, respectively, which is extremely faint and approaches the detection limit. Thus, it is difficult to recover through individual source fitting. 
Previous studies reported much smaller characteristic sizes of $\sim$30 pc for LRDs \cite{Furtak et al.(2024)}, based on gravitationally lensed systems, but did not explicitly separate a potential extended component from the compact PS-dominated core during the fitting. In contrast, our analysis isolates the extended emission through PS subtraction in the stacking procedure and measures the size of this extended component directly.

The size of the stacked extended emission shows a clear wavelength dependence, decreasing from the rest-frame UV to the rest-frame optical. In particular, the intrinsic size is $\sim$392 pc in the rest-frame UV (F115W) and decreases to $\sim$210 pc in the rest-frame optical (F444W). In the rest-frame UV bands, LRDs frequently exhibit more complex and heterogeneous morphologies, including clumpy or multicomponent structures, which probably contribute to the larger effective sizes measured at shorter wavelengths. We explicitly account for the wavelength dependence of the PSF through forward PSF-convolved modeling, and all reported sizes correspond to PSF-deconvolved intrinsic values. Thus, variations in the PSF size across bands do not bias the recovered source sizes. Consistent with this, the stacked extended emission is resolved at all wavelengths, with FWHM values larger than those of the corresponding PSFs (Supplementary Figure~3). In addition, simulation tests under realistic PSF variations and noise levels confirm that intrinsic sizes can be robustly recovered. We therefore conclude that the observed wavelength dependence of the extended emission size is intrinsic and not driven by PSF effects.

\medskip
\noindent {\bf Systematic analysis through simulation and bootstrap tests.} \label{sec:Simulation tests}
We conduct realistic simulation tests to verify the reliability of our fitting results. The simulation consists of two main steps: individual LRD image fitting and stacking of the data$-$PS residuals. 
We use \texttt{lenstronomy}\cite{2021JOSS....6.3283B} to generate 3000 mock LRDs (\sersic\ + PS) using structural parameters inferred from the results of the real sample. The \sersic\ component \reff, \sersic\ index $n$, axis ratio $q$, and the relative positional offsets between the \sersic\ and point source components were randomly drawn within the numerical ranges spanned by the observed LRD population. The total magnitudes were generated around the observed magnitude distribution of the real sample in each band, enabling a modest range of variation. The extended-to-total flux ratios were also drawn to follow the same band-dependent distributions as measured for the real LRD sample.

First, for each mock source, we construct a model image consisting of a \sersic\ component and a central point source. The PSF used to generate the point source and to convolve the \sersic\ model was selected from PSFEx in the corresponding COSMOS-Web tiles (A1--A10 and B1--B10). The mock images were further injected with background noise extracted from empty sky regions and Poisson noise, ensuring that both the sky characteristics and photon statistics match the real data.
All mock images are then analyzed using the same fitting procedure and parameter constraints adopted in the main analysis.

We then randomly select 150 sources to perform stacking tests. For these objects, we obtain the averaged results from stacking and then compare that with the input \sersic\ magnitude and \reff\. The stacking procedure follows the same approach as in the real data; residual images after point source subtraction were aligned using the inferred extended component center positions (center-x and center-y) derived from individual fits, and the residuals were stacked to create mock extended component images, which were then fitted with a single \sersic\ profile to recover the extended component properties. This process was repeated 100 times to assess statistical stability, while ensuring that the simulated realizations adequately cover the range of the measured extended component magnitudes (as shown in Extended Data Figure~3). 

Finally, the simulation results show that the stacking procedure accurately recovers the population-averaged extended component properties. The mean magnitude is reproduced to within $<0.4$ mag, and the mean \reff\ to within $<0.2$ dex. We note, however, that the alignment procedure can introduce a slight underestimate of sizes, especially at shorter wavelengths, but the effect remains within this uncertainty. Overall, this indicates that stacking, unlike individual fits, yields robust and statistically significant constraints on the average extended component properties of LRDs.

In addition, to quantify the intrinsic statistical uncertainties of the LRD population and to assess the impact of sample variance, we performed a bootstrap analysis based on the real dataset. In each realization, we randomly resampled 217 sources with replacement from the original LRD sample, preserving the same sample size as used in the main stacking analysis. The resampled sources were then stacked and fitted using the identical procedure and fitting configuration described in the main text. This process was repeated 400 times to build robust statistical distributions of the recovered extended-component magnitudes and size. The dispersion of these bootstrap realizations provides an empirical estimate of the population-level uncertainties, naturally incorporating the effects of object-to-object variation, noise fluctuations, and fitting degeneracies. These uncertainties are combined in quadrature with those derived from the simulations to obtain the final error estimates reported in Table~\ref{tab:host_props}.

\medskip
\noindent {\bf An independent confirmation of extended emission with a second stacking method.} \label{sec:An independent confirmation of extended emission with a second stacking method}
In addition to the stacking method described in the Methods Section ``Unveils faint extended emission via image stacking'' (hereafter Method 1, or M1), we implemented an alternative stacking and fitting strategy to independently validate our results, referred to as Method 2 (M2). 
This M2 directly stacks all original LRD images without any prior image decomposition, thereby combining the total emission from all sources. Image fitting with PS and \sersic\ components is then performed only once on the stacked image to assess the presence and properties of the extended emission. This alternative workflow provides a complementary validation of the extended component detection that does not rely on object-by-object decomposition.

The key difference between M1 and M2 lies in the order of image decomposition and stacking.
In M1, each individual LRD image is first modeled with a PS + \sersic\ component. The inferred PS component is then subtracted from the original image, leaving the extended emission in the residual. These residual images are subsequently used for stacking.

To ensure consistency between both methods, the selection criteria for the LRD sample used in M2 are identical to those applied in M1, comprising the final set of 217 LRDs. Prior to stacking, we applied interpolation to the individual images and shifted their positions to the pixel centers determined from the M1 fitting. This step is essential because the final stacked image is interpreted as a single point source, so precise alignment of the pre-stacked images is required. The image alignment was performed using the \texttt{shift()} function from the \texttt{SciPy} package. Due to the interpolation process, a slight distortion to the shape of the point source component is inevitable. After alignment, the images were stacked and averaged to produce the final combined LRD image.

We directly compare the sharpness of the stacked LRD images (M2) to that of the PSF and find that the PSF models have significantly smaller FWHM values than the combined LRD images at both short and long wavelengths.
In the FWHM comparison, the stacked M1 images (representing the isolated extended component) show the broadest profiles, while the M2 images exhibit intermediate FWHM values between the PSF and M1
(see Supplementary Figure~3). This systematic broadening, even before explicit point-source subtraction, provides further evidence for the presence of extended emission beyond the central point source.

Next, we perform image fitting on the stacked LRD. First, using the PS-only model, we detect a `ring'-like residual feature similar to that observed in M1, further supporting the existence of extended emission (see Supplementary Figure~4). However, we note that stacking point-like sources inevitably leads to some loss of central sharpness due to small centroid offsets and residual alignment uncertainties. Thus, when fitting with the PS+\sersic\ model, the \sersic\ component tends to absorb nearly all the light. This behavior is likely caused by the interpolation and shifting process distorting the point source profile, which complicates accurate modeling of the point source component in the LRD. Therefore, M2 serves primarily as a qualitative confirmation of extended emission, rather than as a means for precise measurement of the LRD extended emission properties.

\medskip
\noindent {\bf SED fitting analysis based on stellar model.} \label{sec:SED fitting analysis based on stellar model}
Assuming the host emission originated from stellar components, we perform SED fitting to estimate its stellar masses using the photometry measured from four NIRCam filters (F115W, F150W, F277W, and F444W) at $z=6.45$ which straddle the rest-frame 4000 \AA\ break.
We use \texttt{bagpipes}\cite{bagpipes} to perform the fitting, which can generate a set of templates with a range of ages and metallicities.
We adopt a Kroupa IMF, which is consistent with the Chabrier IMF, thus, it allows direct comparison with the local black hole–host galaxy mass relations (e.g., ref\cite{KH13, G20}). Contributions of nebular emission lines from the host galaxies are also included with a uniform log ionization parameter $\log U$ over the range $[-3, -1]$. The star-formation history is modeled with an exponentially declining ($\tau$-model) component. There are four key parameters that define the stellar population: age, $\tau$, stellar metallicity, and dust attenuation $A_V$.
We adopt a uniform prior on the stellar age in the range $[0.01, 0.85]$ Gyr, capped by the age of the Universe at $z=6.45$. The e-folding timescale $\tau$ is allowed to vary between $0.01$ and $0.85$~Gyr, and the stellar mass formed is parameterized by $\log_{10}(M_\ast/M_\odot)$ ranging from $0.1$ to $15$. Stellar metallicity is allowed to vary between $0$ and $2.5Z_\odot$. Dust attenuation is described by a Calzetti law, with $A_V$ ranging from $0$ to $5$ mag.
A random parameter sampling is performed through MCMC to infer the uncertainties of the SED parameters.
As a result, the stellar mass of the stacked host is log$M_{*}={9.02^{+0.20}_{-0.18}}M_\odot$ (see Figure 3), based on a stellar population template.
In addition, we apply the independent SED fitting code~\texttt{gsf}\cite{gsf} and obtain consistent results, with log$M_{*} \sim 9.01^{+0.24}_{-0.18} M_\odot$, further confirming our stellar mass measurement of stellar mass.

\medskip
\noindent {\bf Consistency across redshift bins.}
To further examine whether the wide redshift coverage of our sample affects the stacked results, we repeat the above analysis within finer redshift bins over $5 < z < 9$ with a bin width of $\Delta z = 0.5$. 
We only perform analysis in those redshift bins inside the range $5.5 < z < 7.5$, which contain a sufficient number of data for robust statistical analysis, whereas the other redshift range only contain few data points and could result in large uncertainties.
For each redshift bin, we adopt the same strategy as that of a combined full redshift range to measure the size, magnitude, and stellar mass.
The results of physical parameters are summarized in Supplementary Table~1. We find that both the characteristic sizes and magnitudes remain broadly consistent across redshift bins, showing no significant evolution within the measurement uncertainties.

We note that in some of the higher-redshift bins, the inferred stellar mass is larger than that in the lower-redshift bins. However, this behavior is likely driven by selection effects rather than genuine physical evolution. In particular, the luminosity distributions of the LRD samples in different redshift bins do not show significant differences, implying that at higher redshifts the sample preferentially selects intrinsically more luminous, and hence more massive sources. In addition, the observed variation in size across redshift bins remains within the quoted uncertainties, and therefore does not provide evidence for a statistically significant size evolution.

\medskip
\noindent {\bf SED fitting analysis based on nebular gas model.} \label{sec:SED fitting analysis based on nebular gas model}
The extended emission in LRDs could alternatively originate from nebular gas emission \cite{Chen et al.(2025)}. We therefore test whether this scenario can explain the observed properties.
We adopt the nebular emission templates same as Chen et al.\cite{Chen et al.(2025)} and then perform SED fitting.

The SED fitting is performed using $\chi^2$ minimization, where the photometric uncertainties in each band are used.
The best-fit result adopting template from Chen et al.\cite{Chen et al.(2025)}, can broadly match the observed four-band photometry (see Extended Data Figure~4), and the inferred metallicity is approximately \(0.0002~Z_\odot \).
However, Chen et al.\cite{Chen et al.(2025)} have also reported a detection of a strong [O III] emission line in one case, but our best-fit result reflects that the emission is largely driven by other strong emission lines which have not been spectroscopically detected yet. 
The gas-only model fails to reproduce the observed flux from the spatially extended emission in the F444W band, which is the key observational constraint in this study.
On the other hand, the total $\chi^2$ value of the weighted nebular fit ($\chi^2 \sim 13.48$) is much higher than that of the stellar population model ($\chi^2 \sim 3.03$). Taking all together, our result indicates that the nebular emission alone is unlikely to dominate the extended component, and the stellar population model provides a more physically plausible explanation.

To increase the relative influence of the bluer bands, we additionally test an alternative fitting strategy in which equal weights are assigned to all photometric bands. Under this assumption, the nebular gas model yields a solution that can better reproduce the excess flux in the bluer bands, corresponding to the model labeled (blue bands enhanced) in Extended Data Figure~4. This solution was therefore adopted as the gas component in the composite (stellar + gas) model discussed below, in order to account for the potential contribution of nebular emission to the observed blue-band excess.

Lastly, we have also tested a composite model, i.e., stellar+gas.
The best-fit solution is overwhelmingly dominated by the stellar component ($\sim$93\%), while the gas contribution is consistently driven to its lower bound, even when its allowed fraction is artificially increased. This indicates again that the extended emission is primarily stellar in origin.

\subsection*{Data Availability}
The JWST data are available from the COSMOS-Web survey\url{https://cosmos2025.iap.fr/}. The catalog of LRDs can be accessed at \url{https://github.com/hollisakins/akins24_cw}\cite{Akins et al.(2024)}.

\subsection*{Code Availability}
The following software packages used in this work are publicly available:  

- \texttt{galight}: \url{https://github.com/dartoon/galight} 

- \texttt{bagpipes}: \url{https://github.com/JohannesBuchner/MultiNest} 

- \texttt{gsf}: \url{https://github.com/mtakahiro/gsf.git}

\subsection*{Acknowledgements}
We thank Yuxuan Pang and Shengzhe Wang for fruitful discussions.

\subsection*{Funding Statement}
This work was supported by National Natural Science Foundation of China (Grant No. 12573017). 
X.D. and K.L acknowledge the National Key R\&D Program of China (No. 2024YFC2207400).
M.F. acknowledges funding from the European Union’s Horizon 2020 research and innovation program under the Marie Sklodowska-Curie grant agreement No 101148925.
L.C.H. was supported by the National Science Foundation of China (Grant No. 12233001) and the China Manned Space Program (CMS-CSST-2025-A09).  
K.I. acknowledges support from the National Natural Science Foundation of China (12573015, W2532003), the Beijing Natural Science Foundation (IS25003), and the China Manned Space Program (CMS-CSST-2025-A09).
The Cosmic Dawn Center (DAWN) is funded by the Danish National Research Foundation under grant DNRF140.
S.J. acknowledges the Villum Fonden research grants 37440 and 13160.
K.L acknowledges the National Natural Science Foundation of China (NSFC) No. 12222302.
M.O. is supported by the Japan Society for the Promotion of Science (JSPS) KAKENHI Grant Number 24K22894.  
T.T. is supported by the Japan Society for the Promotion of Science (JSPS) KAKENHI Grant Number JP25KJ0750.
B.Y. is supported by Xiaomi Foundation / Xiaomi Young Talents Program.

\subsection*{Author Contributions Statement}
Y.Z. led the analysis of the LRD decomposition, residual stacking, and simulation tests, and contributed significantly to the preparation of the manuscript.
X.D. conceived the idea of this study, led the SED fitting, and was also actively involved in writing the manuscript.
L.Y. was responsible for the analysis of morphology-related properties and played a significant role in revising the manuscript.
E.L. advised on the sample selection criteria, assisted with the initial stages of the research, and provided detailed suggestions for improving the manuscript's language.
H.A. supplied the COSMOS-Web LRD catalog, supported data processing efforts, and offered guidance on the luminosity limits adopted for source selection.
A.J.B. provided valuable feedback on the manuscript, motivating refinements to several quantitative analyses, and suggested performing checks across multiple redshift bins.
C.M.C. proposed the use of stars as a comparison sample for extended-source detection tests and developed methods for quantitatively comparing the sizes of extended emission in LRDs.
C.C. contributed the gas model used in the fitting.
I.C. reviewed the manuscript's wording and contributed to discussions on the physical implications of stellar mass.
A.F. carefully reviewed the manuscript's language and phrasing and provided suggestions for the design of the multi-redshift-bin tests.
M.F. offered overall feedback on the manuscript and suggested several valuable tests related to the \sersic index.
L.C.H. proposed the nebular gas scenario and improved the stacking methodology, carefully reviewed the manuscript, and offered suggestions for refining the physical descriptions.
K.I. initiated discussions on potential differences among sources at different redshifts, advised on the extended-to-total flux ratio, provided insights into short-wavelength off-center sources, and reviewed Figures 1--4.
S.J. contributed insights into the photometric properties of host galaxies and the relationship between stellar mass and black hole mass.
M.O. closely followed the progress of the initial research and provided valuable guidance on exploration directions and testing strategies.
V.K. offered suggestions for improving the manuscript's wording.
J.D.S. provided overall feedback on the manuscript and contributed to language refinement.
T.S.T. supplied the COSMOS-Web stellar PSF data, carefully reviewed the manuscript, and provided valuable input on wording, physical interpretations, SNR tests, and the implications of the SEDs.
H.Y. suggested an alternative stacking approach (M2) and provided suggestions on the overall study.
J.A.Z. provided insights into the impact of sources with strong extended-to-total flux ratios on the overall results.
A.H., M.K., J.S.K., K.L., N.R., R.M.R. and B.Y. contributed to the discussion of the presented results and the preparation of the manuscript.

\subsection*{Competing Interests Statement} 
The authors declare no competing interests.

\clearpage
\newpage
\section*{Extended Data}

\captionsetup[figure]{name=Extended Data Figure}
\setcounter{figure}{0}

\begin{figure}[h]
\centering
\includegraphics[trim = 0mm 0mm 0mm 0mm, clip, width=\textwidth]{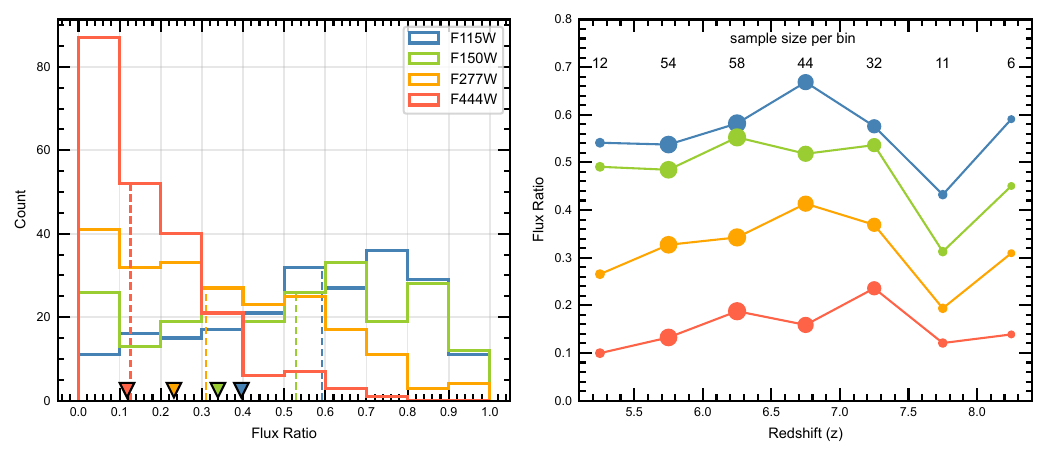}
\caption{\textbf{Distribution of individual LRD extended-to-total flux ratio and the redshift evolution.}
{\it Left:} 
Histogram of the LRD extended-to-total flux ratio. The step histograms represent the distribution of ratio within 10\% intervals. Owing to the low SNR of individual sources, these measurements exhibit large scatter and are shown for illustrative purposes only. The dashed lines mark the median values of individual LRD, while the triangle symbols indicate the extended-to-total flux ratio of the corresponding averaged image in each band using the stacking method which provide the robust population-level constraints adopted in this work.
{\it Right:}
For each redshift bin, the scatter points show the mean extended-to-total flux ratio, with different filters distinguished by the color of the solid lines. The marker size is proportional to the number of sources contributing to each bin. The corresponding source counts for each redshift bin are indicated by the numerical annotations shown at the top of the panel.}
\label{fig:ETR}
\end{figure}

\begin{figure}
	\centering
\includegraphics[trim = 0mm 0mm 0mm 0mm, clip, width=\textwidth]{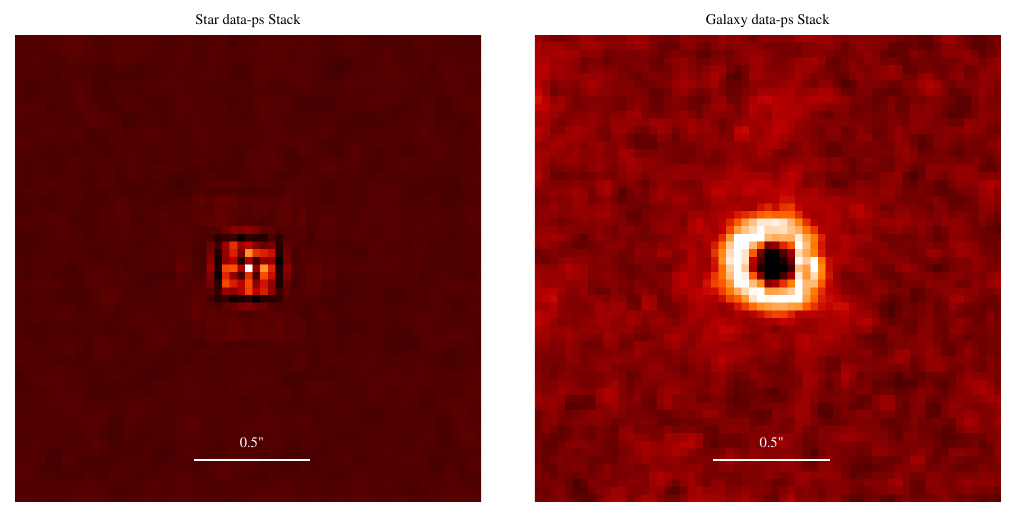}
\caption{\textbf{Stacked residual image of 263 stars and 45 compact galaxies in the COSMOS-Web field after PS-only fitting.} Each star and galaxy was modeled using the corresponding PSF and stacked following the same procedure applied to the LRD sample. 
The residual image of the stacked stars shows no significant extended emission or `ring' features, whereas the stacked compact galaxies exhibit residual structures consistent with those observed in the LRD sample.
}

\label{fig:stack_star}
\end{figure}

\begin{figure}
	\centering
\includegraphics[trim = 0mm 0mm 0mm 0mm, clip, width=\textwidth]{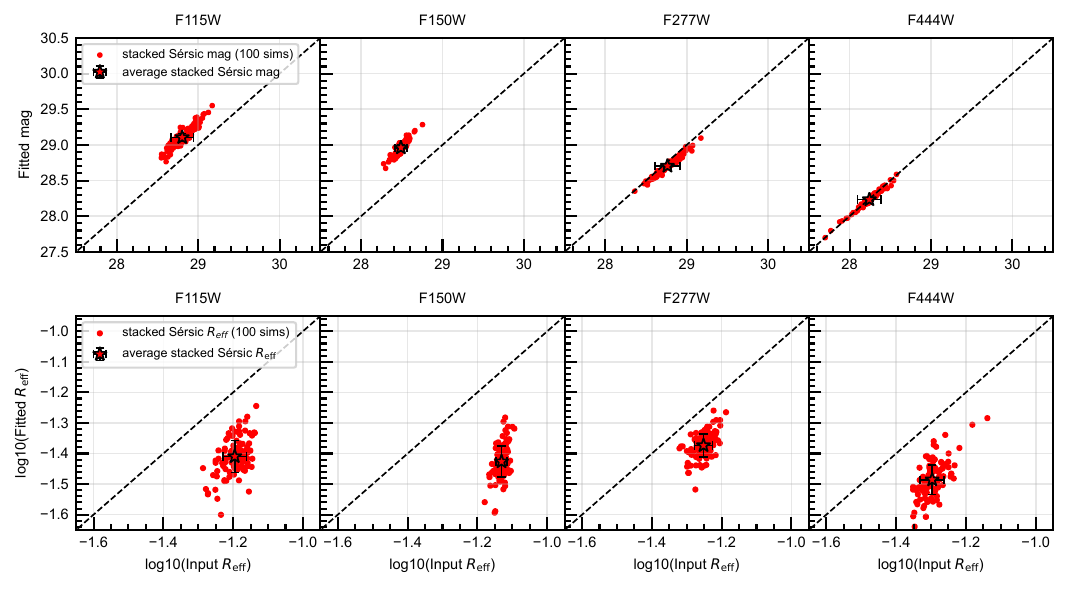}\\
\caption
{\textbf{Simulation results based on 100 realizations.}
{\it Top:} Recovered magnitudes of the simulated host component in each band. The $x$-axis shows the mean input magnitudes of the randomly generated mock \sersic\ components, while the $y$-axis shows the magnitudes recovered from fitting the stacked simulation images. 
{\it Bottom:} Recovered $\log_{10} R_{\rm eff}$ of the simulated host component in each band. The $x$-axis represents the input values, computed as the flux-weighted average effective radius of the randomly generated \sersic\ components, and the $y$-axis shows the fitted results using the stacked host image. 
Red points indicate the measurements from the final stacked images, and red stars mark the averages over 100 realizations.
The dashed black line shows the one-to-one relation. The horizontal error bars denote the standard deviation of the input values, whereas the vertical error bars correspond to the standard deviation of the residuals (fitted$-$input).}

\label{fig:Sim_result}
\end{figure}

\begin{figure}
	\centering
\includegraphics[trim = 0mm 0mm 0mm 0mm, clip, width=\textwidth]{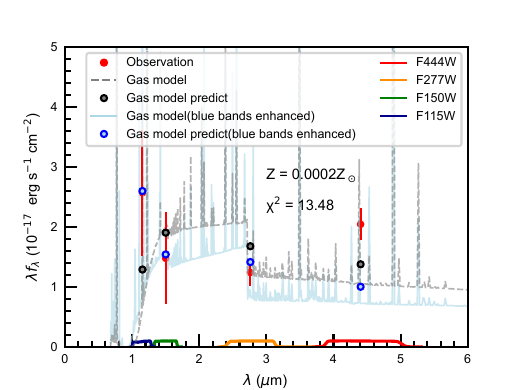}\\

\caption{\textbf{SED fitting of the extended component with a nebular gas model.} Red points with error bars represent stacked fluxes. The gray line shows the model spectrum, and the black circle marks its prediction, with fitting $\chi^2$ indicated. The light-blue line shows the gas model obtained by assigning equal weights to all photometric bands, which enhances the contribution of the bluer bands and better reproduces the excess flux. The blue circle marks the corresponding model prediction. Error bars are derived from simulation-based magnitude uncertainties and converted to uncertainties in the corresponding quantities.}
\label{fig:SED_gas}
\end{figure}

\clearpage

\clearpage
\newpage
\section*{Supplementary Information}

\captionsetup[table]{name=Supplementary Table}
\setcounter{table}{0}

\begin{table}[h]
\centering

\begin{tabular}{ccccccc}
\hline \hline
Redshift bin (Counts) & Filter & Magnitude &  $\langle z \rangle$ & $\log$\reff ($pc$) & $\log(M_*/M_\odot)$ \\
\midrule

5.5--6.0 (54)& F115W & $28.69\pm0.88$ &  5.78 & $2.69\pm0.79$ & $\sim 8.80$ \\
         & F150W & $29.13\pm1.10$ &          & $2.58\pm0.84$     &     \\
         & F277W & $28.51\pm0.38$ &          & $2.46\pm0.36$     &     \\
         & F444W & $27.86\pm0.26$ &          & $2.33\pm0.52$     &     \\
\midrule

6.0--6.5 (58) & F115W & $29.16\pm0.85$ &  6.24 & $2.70\pm0.77$ & $\sim 9.37$ \\
         & F150W & $29.06\pm1.06$ &          & $2.63\pm0.81$     &     \\
         & F277W & $29.12\pm0.37$ &          & $2.44\pm0.35$     &     \\
         & F444W & $27.73\pm0.25$ &          & $2.34\pm0.50$     &     \\
\midrule

6.5--7.0 (44)& F115W & $28.53\pm0.98$ &  6.72 & $2.69\pm0.88$ & $\sim 8.81$ \\
         & F150W & $29.57\pm1.22$ &          & $2.67\pm0.93$    &     \\
         & F277W & $28.62\pm0.42$ &          & $2.51\pm0.40$    &     \\
         & F444W & $28.08\pm0.29$ &          & $2.39\pm0.58$      &     \\
\midrule

7.0--7.5 (32)& F115W & $29.31\pm1.15$ &  7.15 & $2.47\pm1.03$ & $\sim 9.55$ \\
         & F150W & $29.42\pm1.43$ &          & $2.43\pm1.09$     &     \\
         & F277W & $29.07\pm0.49$ &          & $2.32\pm0.47$     &     \\
         & F444W & $27.45\pm0.34$ &          & $2.41\pm0.68$     &     \\
\hline \hline
\end{tabular}
\caption{{\bf Properties of stacked extended emission obtained from data$-$PS across four NIRCam bands in several redshift bins.}
The columns from left to right represent the redshift range (data counts within the range), filter, magnitude, averaged redshift, half-light radius, and stellar mass, respectively. 
$\log$\reff ($pc$) is calculated based on the averaged redshift of each bin, $\langle z \rangle$, and stellar mass is measured from SED fitting.  
The uncertainties of the magnitude and \reff\ are estimated from our simulation-based error analysis and are rescaled according to the number of stacked sources in each redshift bin. Specifically, we apply a scaling factor of $\sqrt{N_{\rm bin}/N_{\rm total}}$, where $N_{\rm bin}$ is the number of sources in the given bin and $N_{\rm total}$ is the total number of stacked LRDs.
}
\label{tab:redshift_bins}

\end{table}

\captionsetup[figure]{name=Supplementary Figure}
\setcounter{figure}{0}

\begin{figure}[h]
\vspace{-0.5cm}
\centering
\includegraphics[trim = 0mm 0mm 0mm 0mm, clip, width=0.45\textwidth]{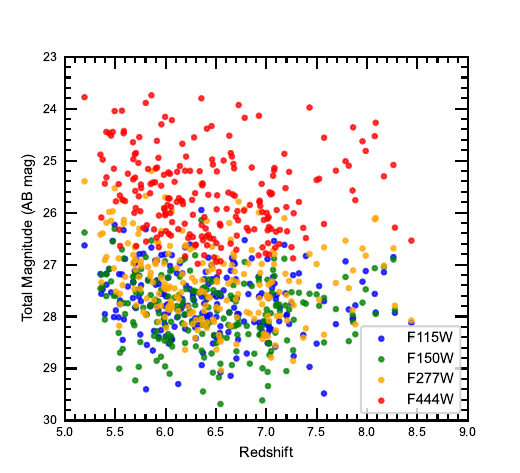}

\caption{\textbf{Distribution of magnitude and redshift for the selected LRD sample in four bands.} The sample includes 217 sources satisfying the criteria \text{F115W} - \text{F150W} $< 0.8$, total magnitude $< 30$, and $\chi^2 < 2$. Red, yellow, green, and blue dots represent F444W, F277W, F150W, and F115W, respectively. The sample spans a redshift range of $z = 5$–9, with source brightness in the short-wavelength bands approaching the detection limit. 
} \label{fig:Samples_mag-z}
\end{figure}

\begin{figure}
	\centering
\includegraphics[trim = 0mm 0mm 0mm 0mm, clip, width=\textwidth]{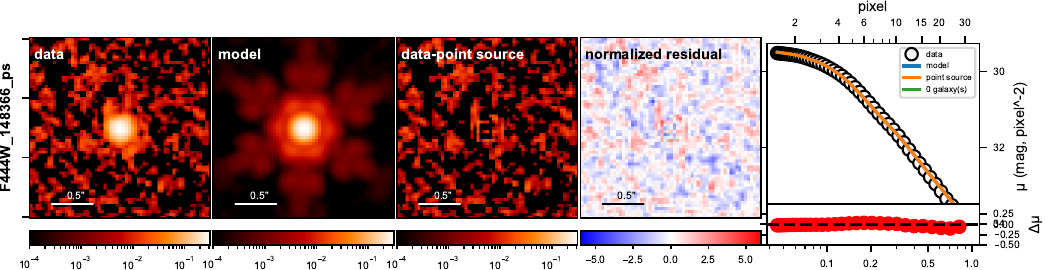}\\
\includegraphics[trim = 0mm 0mm 0mm 0mm, clip, width=\textwidth]{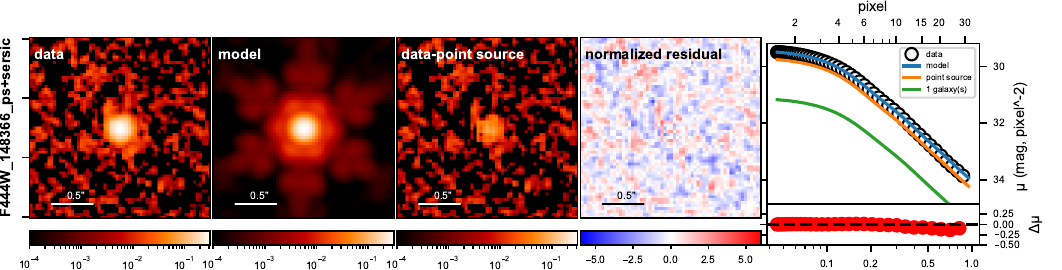}\\
\caption{\textbf{Example of PS-only and PS+\sersic\ fitting on an individual LRD.} 
{\it Top:} Results from PS-only fitting. Panels from left to right: observed image (data), best-fit point source model (PS), residual image after subtracting the PS component (data$-$PS), normalized residual image, and the corresponding 1D surface brightness profile. 
{\it Bottom:} Results from PS+\sersic\ fitting. Panels from left to right: observed image (data), best-fit composite model (PS+\sersic), residual image after subtracting the PS component (data$-$PS), normalized residual image, and the corresponding 1D surface brightness profile.
Both fitting strategies yield acceptable solutions for individual sources. The presence of extended emission can therefore only be robustly confirmed through the stacking analysis.
}
\label{fig:individual_fit_example}
\end{figure}

\begin{figure}
	\centering
\includegraphics[trim = 0mm 0mm 0mm 0mm, clip, width=0.5\textwidth]{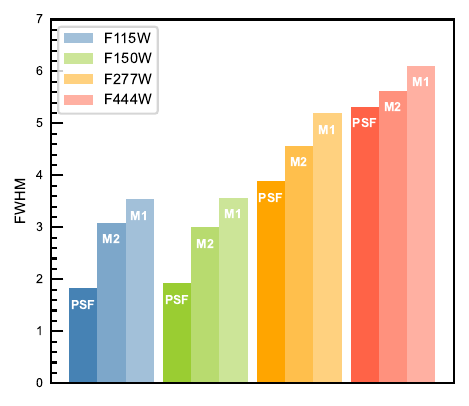}\\
\caption{\textbf {FWHM for PSF, Method-1 and Method-2.} The figure shows the PSF FWHM in the four NIRCam bands, together with the FWHM of the stacked extended emission after point source subtraction (denoted as M1) and the FWHM of the directly stacked LRD images without prior decomposition (denoted as M2).}
\label{fig:FWHM}
\end{figure}

\begin{figure}
	\centering
\includegraphics[trim = 0mm 0mm 0mm 0mm, clip, width=\textwidth]{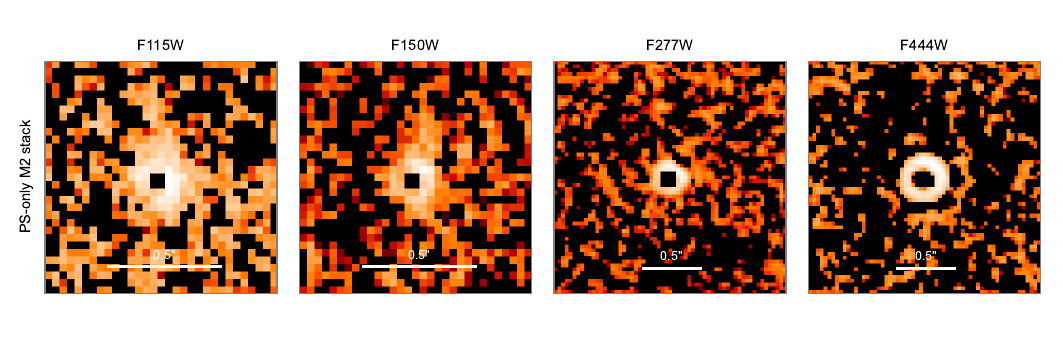}\\
\caption{\textbf{Residual image after performing PS-only fitting on the Method-2 stacked image.} Similar to the residuals obtained by stacking individual data$-$PS images, this approach reveals significant leftover emission after PS-only fitting. The central region exhibits signs of over-subtraction, indicating that a single point source model is insufficient to account for the full light distribution in the outer region.
\label{fig:StackM2psonly}}
\end{figure}

\end{document}